\documentclass[fleqn,10pt]{wlscirep}
\usepackage[utf8]{inputenc}
\usepackage[T1]{fontenc}
\title{Impact of Stain Variation and Color Normalization for Prognostic Predictions in Pathology}

\author[1]{Siyu (Steven) Lin}
\author[1]{Haowen Zhou}
\author[2]{Richard J. Cote}
\author[2]{Mark Watson}
\author[3]{Ramaswamy Govindan}
\author[1,*]{Changhuei Yang}
\affil[1]{California Institute of Technology, Department of Electrical Engineering, Pasadena CA 91125 USA}
\affil[2]{Washington University School of Medicine, Department of Pathology and Immunology, St. Louis MO 63110 USA}
\affil[3]{Washington University School of Medicine, Department of Medicine, St. Louis MO 63110 USA}

\affil[*]{chyang@caltech.edu}

\begin{abstract}
In recent years, deep neural networks (DNNs) have demonstrated remarkable performance in pathology applications, potentially even outperforming expert pathologists due to their ability to learn subtle features from large datasets. One complication in preparing digital pathology datasets for DNN tasks is variation in tinctorial qualities. A common way to address this is to perform stain normalization on the images. In this study, we show that a well-trained DNN model trained on one batch of histological slides failed to generalize to another batch prepared at a different time from the same tissue blocks, even when stain normalization methods were applied. This study used sample data from a previously reported DNN that was able to identify patients with early stage non-small cell lung cancer (NSCLC) whose tumors did and did not metastasize, with high accuracy, based on training and then testing of digital images from H\&E stained primary tumor tissue sections processed at the same time. In this study we obtained a new series of histologic slides from the adjacent recuts of same tissue blocks processed in the same lab but at a different time. We found that the DNN trained on the either batch of slides/images was unable to generalize and failed to predict progression in the other batch of slides/images (AUC\textsubscript{cross-batch} = 0.52 - 0.53 compared to AUC\textsubscript{same-batch} = 0.74 - 0.81). The failure to generalize did not improve even when the tinctorial difference correction were made through either traditional color-tuning or stain normalization with the help of a Cycle Generative Adversarial Network (CycleGAN) process. This highlights the need to develop an entirely new way to process and collect consistent microscopy images from histologic slides that can be used to both train and allow for the general application of predictive DNN algorithms.
\end{abstract}
\begin{document}

\flushbottom
\maketitle
%
%
\thispagestyle{empty}


\section*{Introduction}
The past decade has witnessed wide applications of deep neural network (DNN) models in biomedical and digital pathological analysis due to the growing complexity and training techniques of DNN models. There are increasing examples of the ability of DNN-based methods succeeding in tasks that are simply beyond the scope of even expert pathologists. For example, Bychkov et al. demonstrated that their trained DNN model for analyzing H\&E stained colorectal cancer tumor tissue microarray digital images was able to classify patient risk more accurately than human pathologists\cite{bychkov_deep_2018}. Hekler et al. showed that their trained DNN outperformed 11 pathologists in the classification of histopathological images between benign nevi and malignant melanoma\cite{hekler_deep_2019} and Tschandl et al. demonstrated that DNNs consistently outperform expert pathologists in classifying pigmented skin lesions\cite{tschandl_comparison_2019}.  Bejnordi, et al. showed that trained deep learning models outperformed expert pathologist in simulated time-constrained settings for detecting lymph node metastases in breast cancer tissue sections\cite{ehteshami_bejnordi_stain_2016}. We recently completed a pilot study that showed that DNN can be trained to predict the subsequent development of brain metastases in patients with early stage NSCLC based on images obtained from Hematoxylin-and-Eosin (H\&E) stained slides of their primary tumors\cite{zhou_aiguided_2024} – a task that human pathologists can do little or no better than a random guess. This clearly demonstrates that DNN training can detect subtle features in these images that are simply not recognizable by even expert pathologists.

However, DNN’s capability to detect and use these subtle features is tempered by DNN’s vulnerability to fixate on extraneous variations, in particular variations in tinctorial qualities that are a well-recognized and inherent aspect of the staining of histologic tissue sections with vital stains, a problem that is amplified when training is done on small, homogeneous, well qualified data sets that would otherwise be ideal for DNN training\cite{niazi_digital_2019,bilgin_digitally_2012}. The most straightforward way to force the model to ignore these extraneous variations in staining qualities is to train the DNN with an enormous and varied amount of data (i.e., images) such that, ideally, the model will see these different variations and eventually learn to ignore them\cite{strom_artificial_2020,pantanowitz_artificial_2020}. This method is well suited for everyday tasks such as ImageNet classification networks where there are a lot of public-domain images available for training. However, this ‘big data’ AI training approach is not well suited for histopathology prediction applications. Prospective large histopathology image data sets with known disease outcomes are difficult and costly to assemble, especially for rare diseases. This is especially true as the “best” source of histologic images that would be useful for training a potential predictive algorithm is from a clinical trial with known treatment and outcomes, which is inherently limited or simply do not exist.  Even where the data exists and is accessible, the data format and collection methodology can vary widely across the data collection sites. An alternate strategy to this ‘big data’ approach is to curate or modify the data set to reduce variations prior to DNN training and usage. This approach is well-suited for pathology applications, as far more control can be exerted over the sample and data preparation process in pathology than in most other image-based applications. However, this approach has the disadvantage of being reliant on a smaller data set, more prone to being unable to factor in variations in alternative histology preparations, and thus far less generalizable. 

\begin{figure}[ht!]
\centering
\includegraphics[width=\linewidth]{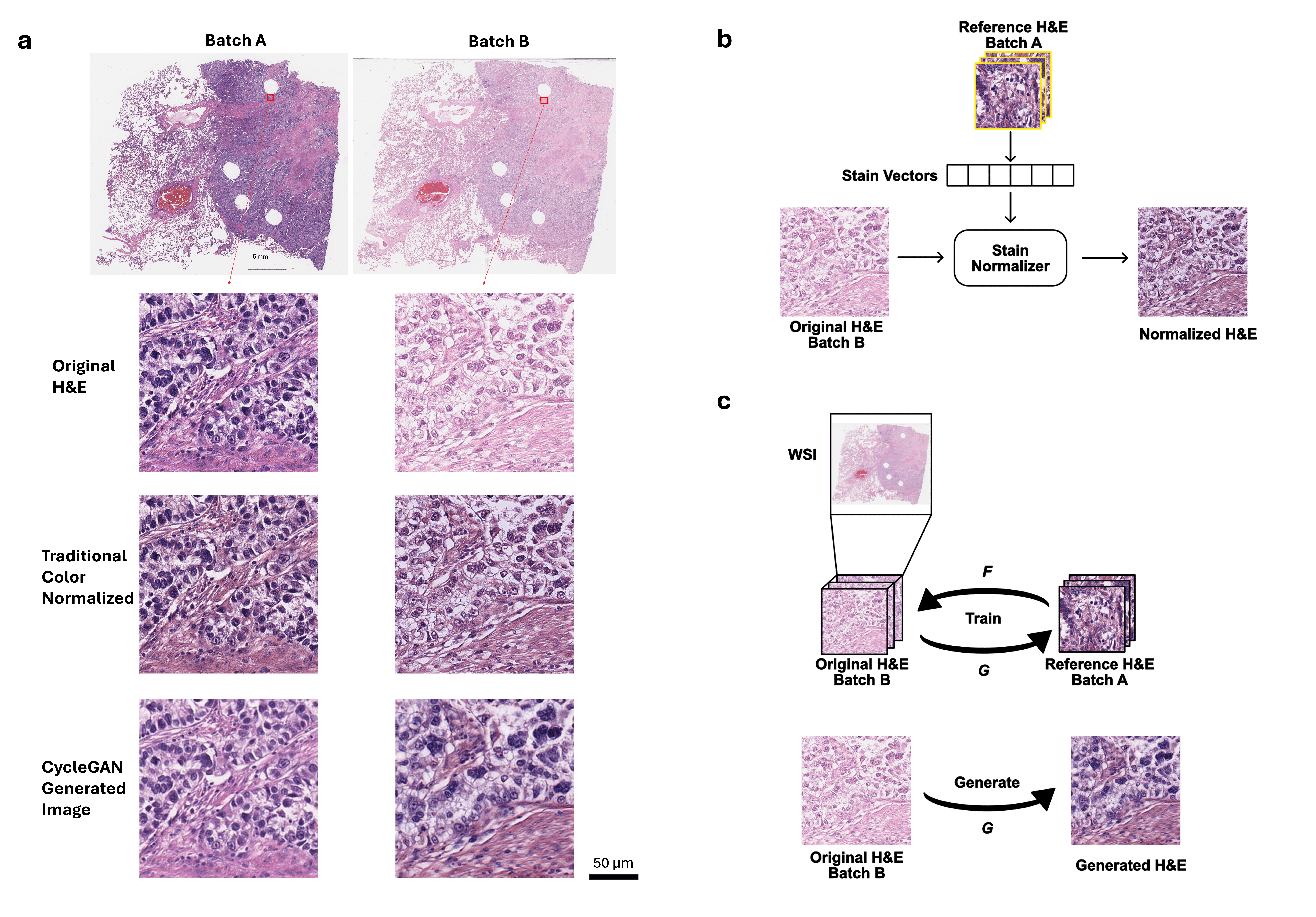}
\caption{a) Stain variations in H\&E-stained adjacent non-small cell lung cancer (NSCLC) histology slides. The left and right slides (top row) are adjacent cuts from the same tissue block from the same patient. Zoom-in regions show original H\&E-stained images (top row) as well as their color-normalized images with a traditional method (middle row) and CycleGAN generative method (bottom row) demonstrating the effects of different color-normalization schemes.  b) Illustration of the traditional method for color normalization of an image where batch B images are normalized to batch A images. c) Illustration of a generative method for color normalization via CycleGAN.}
\label{fig:Fig1}
\end{figure}

Towards this end, color normalization of digital pathology data is an active area of research. One common image processing-based normalization method that performs stain normalization\cite{vahadane_structure-preserving_2016} is purely based on image analysis so that it can learn the stain strengths from one sample image or a collection of example images and normalize all other images towards this staining (Fig. 1b). This method performs sparse non-negative matrix factorization to separate different staining in the source and target images and then color-normalize the images while preserving the structures in the images\cite{vahadane_structure-preserving_2016}. The limitation with such method is that it requires the sample images to be representative of the whole dataset in terms of the staining and cellular contents. More recent machine learning-based methods allow the algorithm to take morphological structures and cellular context into account\cite{ehteshami_bejnordi_stain_2016,khan_nonlinear_2014}. Normalization can also be achieved through a generative model like a Cycle Generative Adversarial Network (CycleGAN) which has demonstrated success relative to other normalization methods\cite{shen_federated_2023}. After training, such generative networks can take an image from one color space and transfer it into a different color space or they can paint an unstained image with pseudostainings\cite{kang_stainnet_2021,rivenson_phasestain_2019}. Color normalization can be performed this way by projecting all images to one single color space (Fig. 1c) specified by a set of reference images. CycleGAN based method allows the potential change of cell morphology in the normalized image whereas the traditional method preserves the structures in the images.

Fig. 1a shows a pair of example images of non-small cell lung cancer (NSCLC) tissue slides that are adjacent cuts from the same tissue block from the same patient. They were cut and stained in the same laboratory but at a different time. Since the two cuts were next to each other, one would expect the cellular content to be similar. However, in the original (batch A) H\&E images, the staining in the slide was heavier/darker (more basophilic) than in the second set of stains on the adjacent cut (batch B), leading to a contrast difference between the image pair. The traditional color normalization method\cite{vahadane_structure-preserving_2016} was able to reduce the contrast difference somewhat, but the remaining differences are apparent, and one can easily appreciate that batch A is still darker than batch B, a difference that can be especially appreciated in the nuclei, where nuclear detail in batch B is more apparent (note easily identifiable nucleoli) compared to batch A, where nuclear detail is obscured due to the darker (more basophilic) staining (Fig 1a). Using a CycleGAN-base method, the tinctorial qualities between batch A and B appeared more similar; however, in this case, note that the cellular morphology between the batches has altered, most notably in the nuclei, where the nuclei in batch B appear larger and more pleomorphic than in the images for batch A.

In this study, we have set out to determine the impact of staining variation on the generalizability of a predictive DNN algorithm. We have explored key issues inherent in DNN driven analysis, including whether predictive algorithms, especially those derived from smaller and more homogeneous data sets, whether advanced color normalization schemes can address the inherent and pervasive impact of stain variability, and if there might be alternative approaches to this important barrier to DNN training and general use.

\section*{Materials and Methods}
\subsection*{Patient Cohort and Whole-Slide Imaging}

The cohort of patients were all diagnosed with stage I-IV NSCLC all diagnosed and treated at Washington University School of Medicine with long-term follow-up (> 5 years or until metastasis)\cite{zhou_aiguided_2024}. This is the same cohort as was used in Ref.\cite{zhou_aiguided_2024}. A total of 198 patients was included in the study and one representative block of tumor tissue from each patient was used to create a fresh H\&E slide which was then scanned at 40× magnification with an Aperio/Leica AT2 slide scanner (Leica Biosystems, Deer Park, IL, USA). The original cuts, denoted batch A, and the recuts, denoted batch B, were prepared in the same laboratory at a different time (separated by ~ 8 months). Each batch was prepared within a relatively short time window. All slides were initially subject to blind review to assess tumor adequacy by an expert pathologist and annotated for ROI by circling an approximate contour of the primary tumor, including the entirety of the tumor microenvironment. Forty-four cases were disqualified as being non-representative or insufficient for adequate evaluation or had a missing recut slide. The remaining 154 cases were used for this study with 63 of them developing metastasis (Met+) to the central nervous system and 91 with no recurrence (Met-). The median time to progression or the follow-up time of these cohorts was 12.2 and 106 months, respectively. Detail information of the stages and the histology of the tumors are summarized in Table 1. The DNN model is blind to the clinical parameters such as stage and histology. 

\begin{table}[!t]
\centering
\caption{Clinical characteristics of the cohort in this study}
\includegraphics[width=\linewidth]{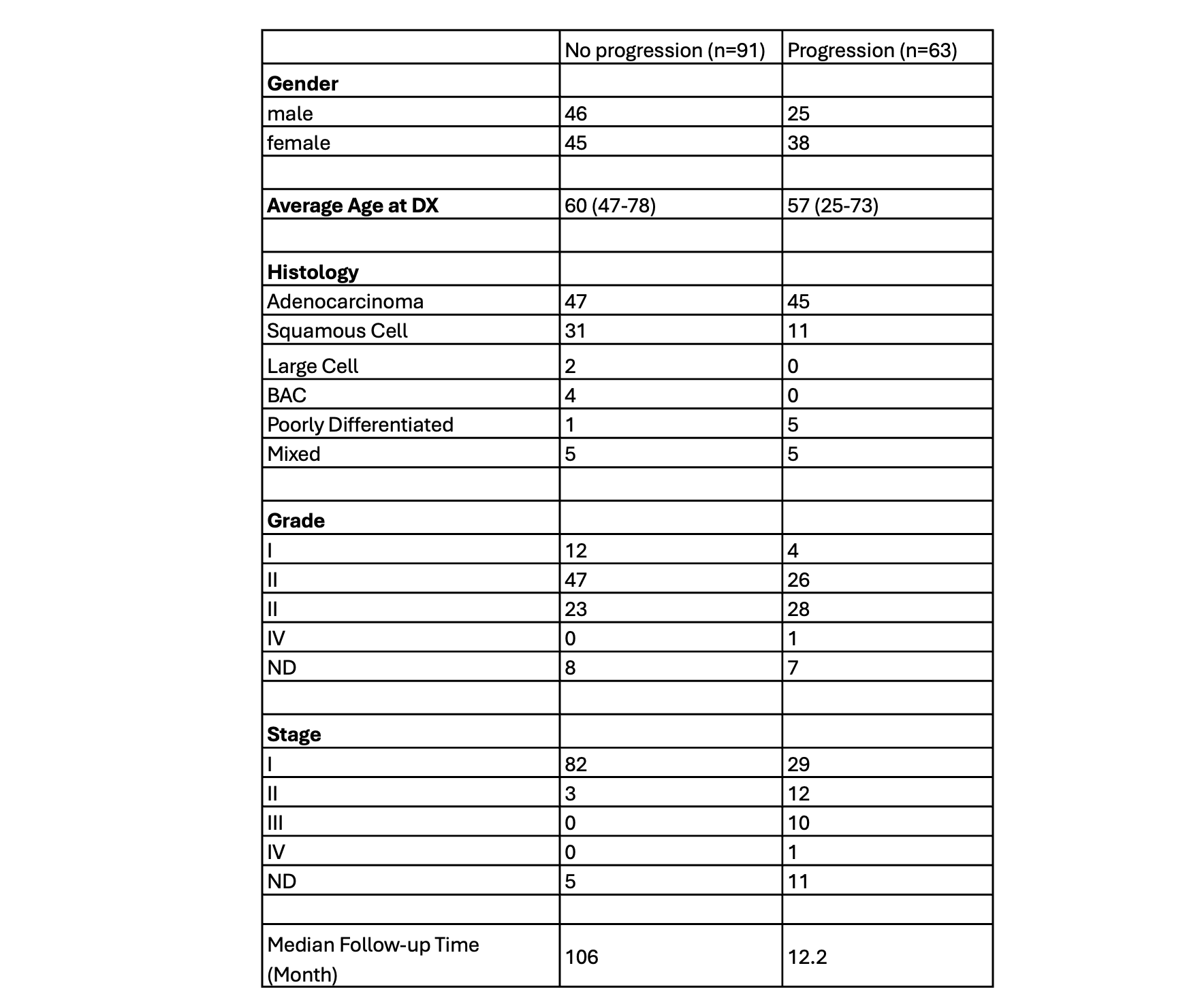}
\label{tab:Tab1}
\end{table}

\subsection*{Data Preprocessing}
The data preprocessing procedure used for this study is similar to the one in Ref.\cite{zhou_aiguided_2024}. The Otsu thresholding method\cite{otsu_threshold_1979} was used on the annotated regions to further exclude the empty areas for each whole-slide image (WSI) and to form the ROI on each slide. Then 1,000 image tiles were randomly sampled from each WSI, each with 256 x 256 pixels or 130 x 130 $\mu$m\textsuperscript{2} under 20x magnification, digitally down-sampled from a 512 x 512 pixels 40x magnification image. In the training process, random rotations, random flips and random crops to 224 x 224 pixels were performed as data augmentation. 

\subsection*{Deep learning study design}
The DNNs were based on the ResNet-18 convolutional neural network pretrained on the ImageNet dataset\cite{he_deep_2015-1}. The models were initialized with the pretrained weights, and all model layers were unfrozen during the training process. A linear layer was attached at the end of the model that followed by a sigmoid activation function that generated a normalized prediction score from each individual tiles. The prediction scores were then supervised with the known Met+ or Met- outcomes. Finally, the progression risk prediction of a WSI was given by the median of the prediction scores of all the individual image tiles associated with the slide.

Since the total number of patients in the study was only 154, to avoid potential bias in the testing set selection from a single experiment, we used a 3-fold experiment format with different training–testing splits where the training set and testing set will be comprised of data from different patients. No color normalization was performed in these control experiments (no color correction involved).  The entire cohort of patients was randomized and numbered from 1 to 154. The randomized patient sequence was used to divide the cohort into a training/validation set (n = 118; Met+ n = 45, Met- n = 73) and a testing set (n = 36; Met+ n = 18, Met- n = 18) in each experiment fold. Specifically, in the first fold of this experiment, the model will be trained on original cuts (batch A) data of patient 1-118 and then tested on batch A and adjacent recuts (batch B) data of patient 119-154. Fig. 2 illustrates this particular experiment. In the other folds of this experiment, we used batch A and batch B data subjects 83-118 and 47-82 as their respective testing set and trained on batch A data of the rest of the subjects. 

A similar experiment was then repeated with the traditional color normalization method. We extracted 100 random image tiles for each slide in the training set. As shown in Fig 1b, the staining vectors of each selected tiles were computed based on sparse non-negative matrix factorization outlined by Vahadane et al.\cite{vahadane_structure-preserving_2016}, and the mean of the staining vectors were used to perform structural-preserving normalization for every image tile in the testing set. The experiments followed the same format as the control experiments. Fig. 3 shows an example data flow chart and the placement of the color normalization step in the process. 

Finally, the experiments were repeated with generative color normalization. For every slide from the testing set, regardless of whether it was from batch A or batch B, we trained a CycleGAN with unpaired data that alternatively project between the images from the testing slide and images from the slides in the entire training set until the generated images from the testing slide is indistinguishable from the training set images to the model discriminator. We then fed the generated images into the trained models to test the model performance. Once again, the experiments followed the same format as the control experiments. Fig. 4 shows an example data flow chart and the placement of the CycleGAN color normalization step in the process. 

\begin{figure}[!h]
\centering
\includegraphics[width=\linewidth]{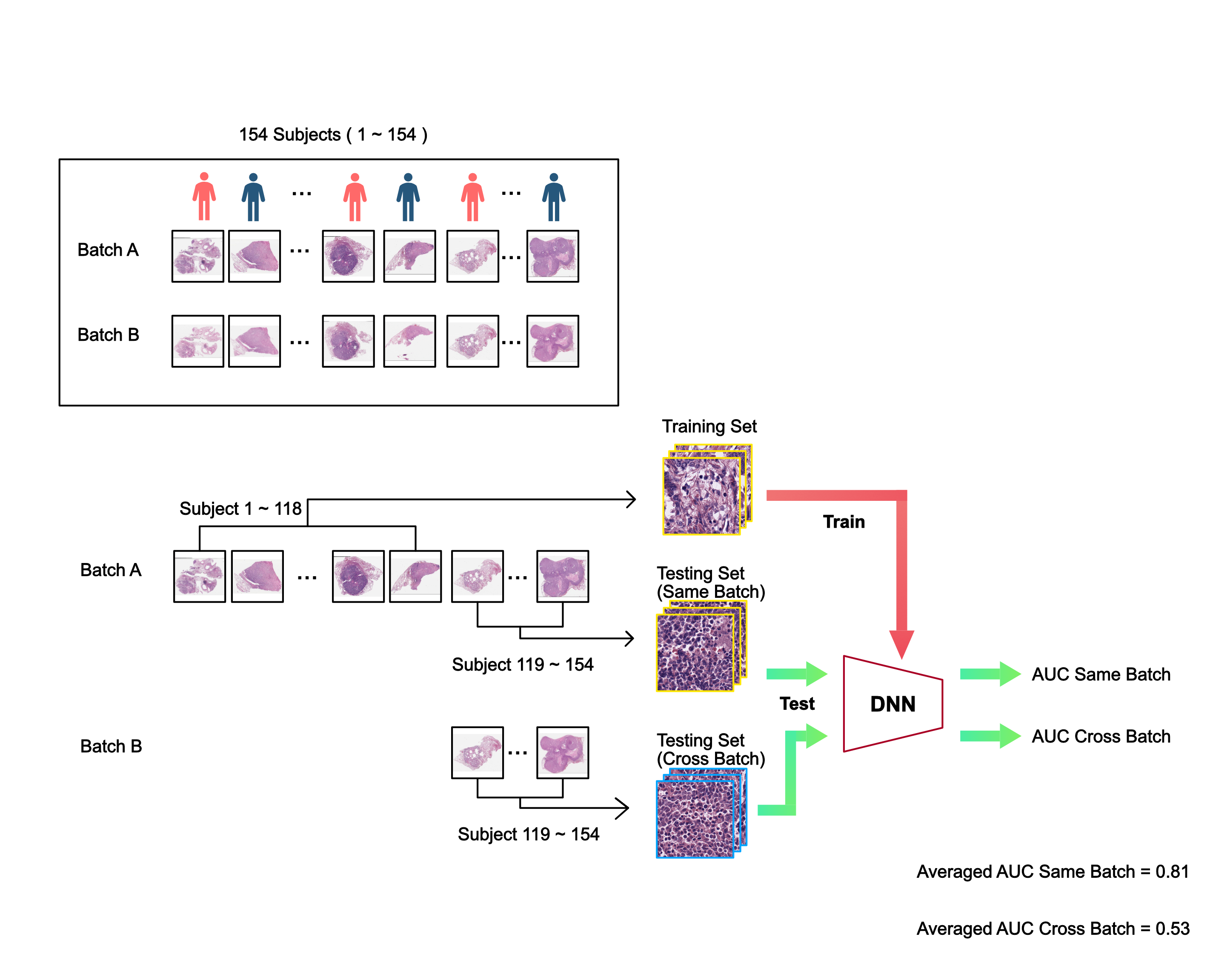}
\caption{Standard training and testing flow chart. The cohort consists of 154 subjects, randomized and labeled with indices 1-154. The subject’s biopsy histopathology slides were prepped and scanned on two separate occasions, resulting in two separate pathology image sets denoted by batch A and batch B. In the example illustrated above, we reserved slides of subject 119-154 of batch A as the same-batch testing set, and the rest of the batch A as the training set. The slides of the same subject (subject 119-154) from batch B served as the cross-batch testing set. The trained DNN was then be used to analyze the same-batch testing set slides to yield a AUC\textsubscript{same-batch} value. We performed the same DNN classification on the cross-batch testing set slides to yield a AUC\textsubscript{cross-batch} value. This whole training-and-analysis process was repeated three-fold by using different subjects as the test subjects. Specifically, fold-2 would use subject 83-118 as test subjects and fold-3 would use subject 47-82 as test subjects. The average AUC\textsubscript{same-batch} and AUC\textsubscript{cross-batch} from the 3-fold experiment are reported above.}
\label{fig:Fig2}
\end{figure}

\subsection*{Statistical Analysis}
To assess the effectiveness of our DL-based classifier in predicting progression risk, the area under the receiver operating characteristic (ROC) curve (AUC) was calculated to provide a measure of the overall performance of the model. To compare the model outputs with the ground-truth clinical progression outcomes, we binarized the model prediction scores and reported the accuracy metric. p values were calculated to show the performances difference of the model tested with color normalized images compared with the model tested with original H\&E images.

\section*{Results}

The above-described study design allowed us to generate an AUC score for when a DNN was trained on training data from a batch and tested on a set-aside collection of test slides from the same batch. We labelled this AUC score as AUC\textsubscript{same-batch}. We can also generate a different AUC score - AUC\textsubscript{cross-batch}, for when a DNN was tested on corresponding slides from the same subjects but from the other batch. As we performed each experiment in 3-fold, we can compute the average of these AUC scores. As noted in Fig. 2, when the model was trained on training data from batch A and tested on set-aside data from batch A, the model was able to successfully predict metastatic outcome with an average AUC\textsubscript{same-batch} of 0.81. This corresponded well with the results reported in Ref.\cite{zhou_aiguided_2024}. (We noted that the results are not identical because some slides used in Ref.\cite{zhou_aiguided_2024} were excluded due to lack of corresponding adjacent cuts in batch B, and traditional color normalization was performed in Ref.\cite{zhou_aiguided_2024}.) The predictive power of this model was statistically significant compared to a no-prediction-value null hypothesis with a p-value < 0.0001. Interestingly, the model trained on training data from batch A failed to generalize when tested on testing set data from batch B yielding an average AUC\textsubscript{cross-batch} scores of 0.53 with p-value > 0.05 compared to a no-prediction-value null hypothesis. In summary, in a control experiment with no color normalization, a DNN can be trained to make statistically meaningful prediction when the training set and testing set data originated from the same batch (but did not contain overlapping slides), but the same trained DNN failed to generalize to the other batch. 

\begin{figure}[!t]
\centering
\includegraphics[width=\linewidth]{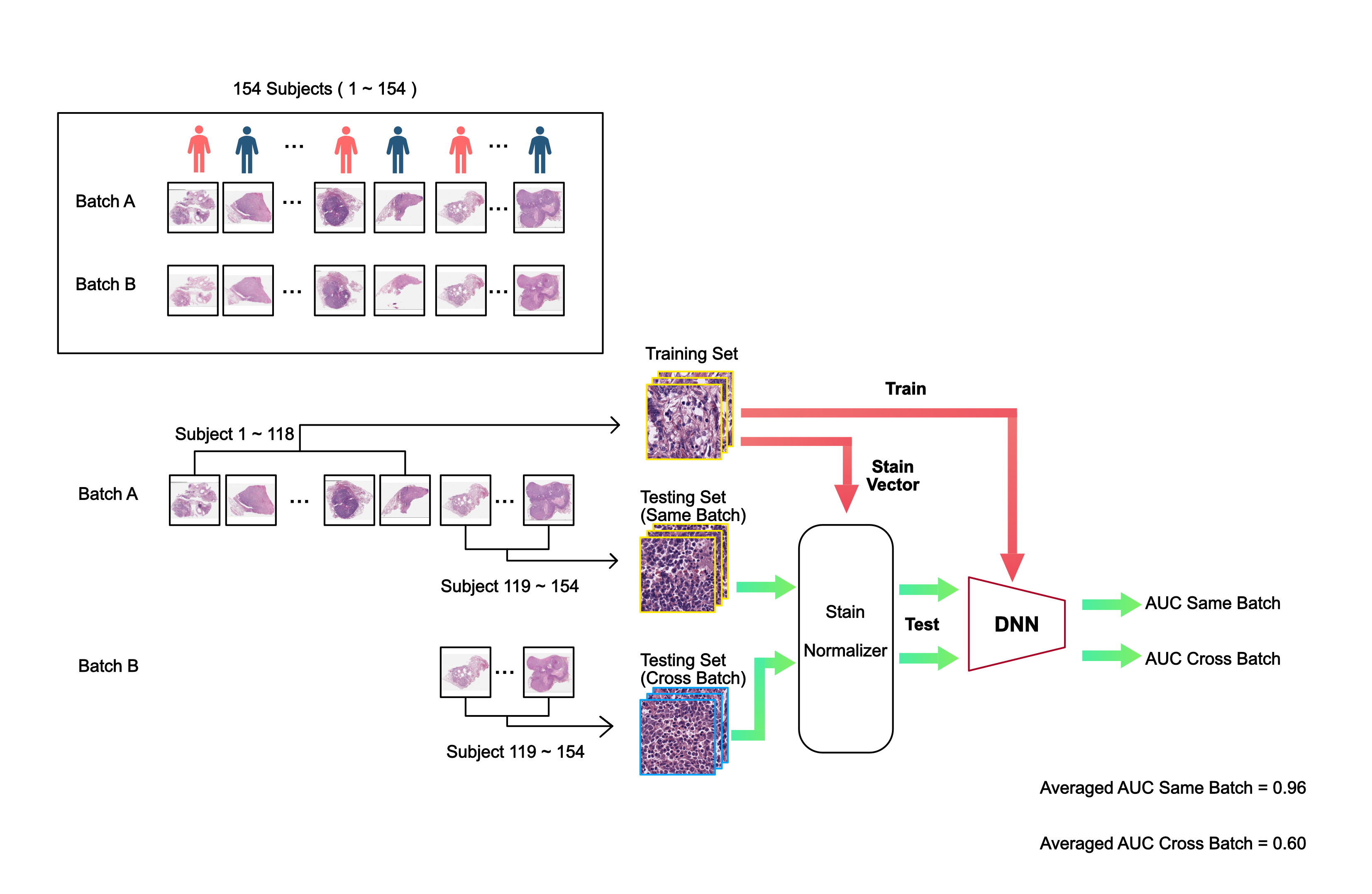}
\caption{Traditional color normalization training and testing flow chart. The processing sequence is similar to that for the standard training and testing flow chart. Here, we performed traditional color normalization on the resting set images prior to feeding them through the trained DNN for classification.}
\label{fig:Fig3}
\end{figure}

The traditional color normalization experiment (see Fig. 3) yielded an averaged AUC\textsubscript{same-batch} of 0.96. However, the model trained on training data from batch A failed to generalize when tested on testing set data from batch B yielding an average AUC\textsubscript{cross-batch} scores of 0.60 with p-value > 0.05 compared to a no-prediction-value null hypothesis.

\begin{figure}[!t]
\centering
\includegraphics[width=\linewidth]{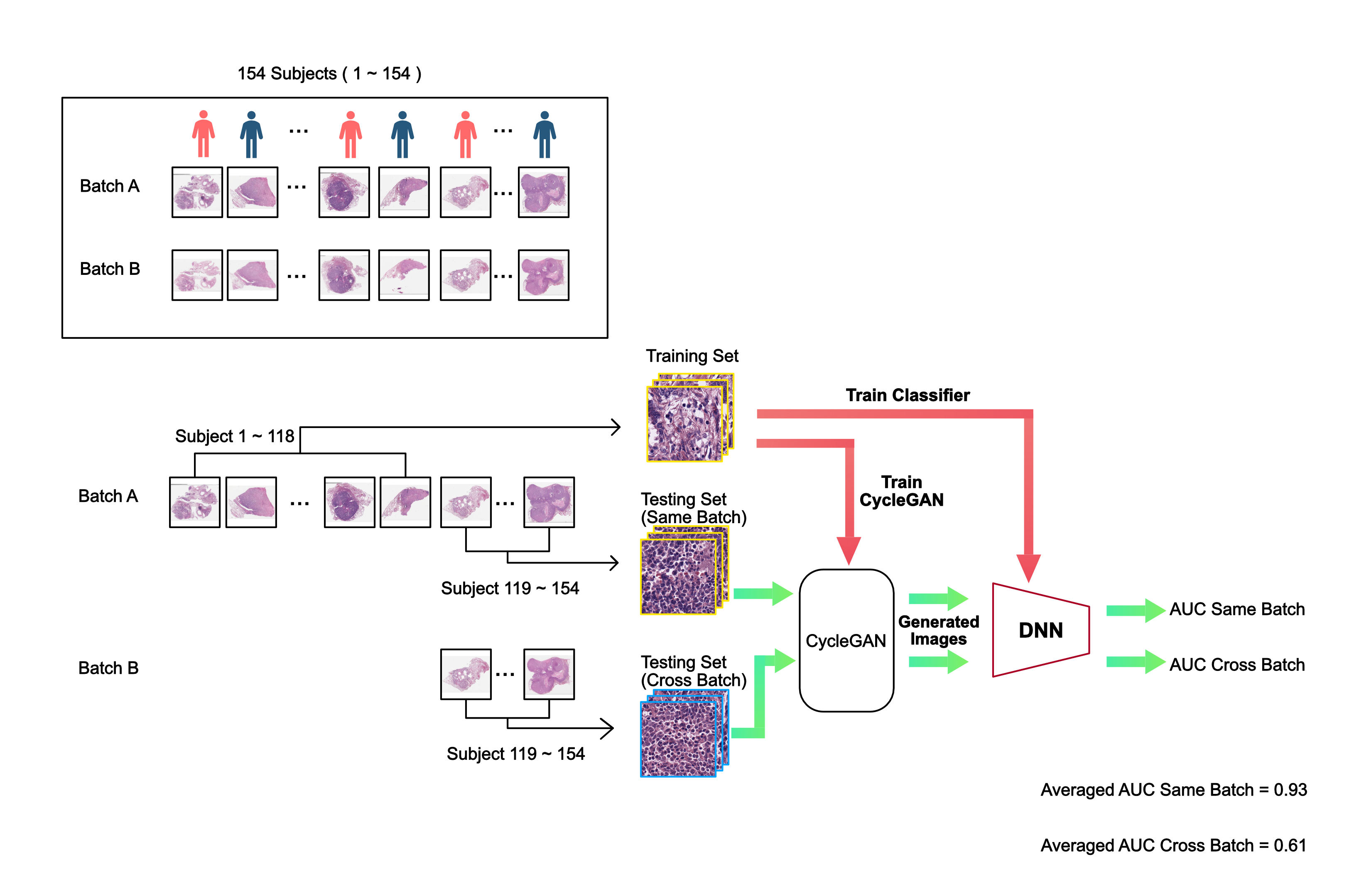}
\caption{Generative color normalization training and testing flow chart. The processing sequence is similar to that for the standard training and testing flow chart. Here, we performed traditional color normalization on the resting set images prior to feeding them through the trained DNN for classification.}
\label{fig:Fig4}
\end{figure}
In the final experiment where CycleGAN color normalization was employed, the DNN achieved an averaged AUC\textsubscript{same-batch} of 0.93. In addition, the model trained on training data from batch A failed to generalize when tested on testing set data from batch B yielding an average AUC\textsubscript{cross-batch} scores of 0.61 with p-value > 0.05 compared to a no-prediction-value null hypothesis.

In all the experiments described above, the models were trained with batch A data and evaluated on batch A and batch B data of the patients in the testing sets. We can also swap the role of Batch A and B and repeat all of the experiments, with the only difference being that we train the models on batch B instead and we color normalize the testing images towards the training set of the batch B data. All the results are summarized in Table 2. When we train on batch B, we see similar trends in that when the model was tested cross-batch (on batch A), the model failed to generalize. 

\begin{table}[ht]
\centering
\caption{Results summary of all experiments where we train the models on either batch and test the models within the same-batch or cross-batch. We perform hypothesis testing where the null hypothesis states that the model testing accuracy with color normalization is the same as using just the original H\&E images and the alternative hypothesis as the model testing performance with color normalization is better than with the original H\&E images. p-values are indicated in the parenthesis.}
\includegraphics[width=\linewidth]{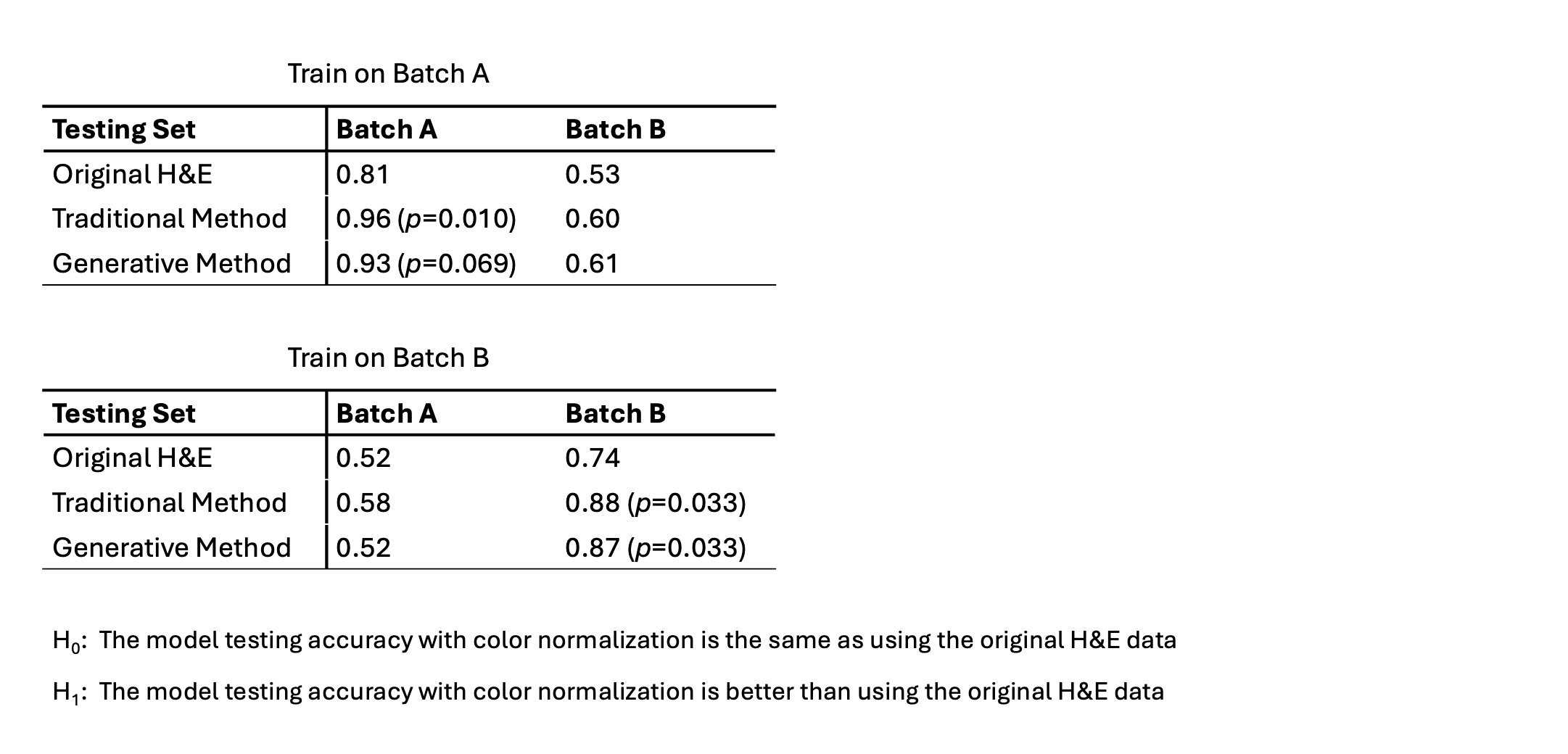}
\label{tab:Tab2}
\end{table}

With regards to color normalization impact on DNN’s performance on same batch data, the slightly improved AUC\textsubscript{same-batch} scores suggest there may be some improvements. However, when checking the associated p-value with a Bonferroni multi-comparison corrected alpha of 0.05/4, we found that in all case except one, the increased AUC\textsubscript{same-batch} did not rise above statistical significance. The sole exception is that the increased AUC\textsubscript{same-batch} for the experiment where traditional color normalization was used for the DNN trained on training set data from batch A and tested on testing set data from batch A. That case gave a p-value of 0.010 which is slightly lower than the Bonferroni corrected alpha of 0.0125. However, the p-value is so close to the alpha that we would caution against overreading into the statistical significance of that result.

The failure of color normalization to help DNN generalize in this case study points to the fact that current color normalization methods are not capable of fully mapping subtle stain variations with sufficient fidelity between data sets with different staining contrasts. This lack of fidelity is in fact visible to the expert human eyes, as further seen in Fig 5. We note that the tinctorial qualities of Batch A and B are different, with greater basophilia in batch A.  Using Batch A H\&E images as the baseline for normalization, we can appreciate that the traditional color normalization corrects this to some extent, although there are clear differences between the Batch A and normalized Batch B; for example, note that the nuclei in Batch A are more basophilic, and there is greater nuclear detail in normalized Batch B, such as the clear presence of nucleoli. The CycleGAN generated Batch B images appear to have somewhat more consistent tinctorial qualities with Batch A (baseline); however, the CycleGAN process appears to have resulted in distinct morphological changes, particularly notable in the nuclei, which appear more basophilic and pleomorphic compared to Batch A baseline.

\begin{figure}[h!]
\centering
\includegraphics[width=\linewidth]{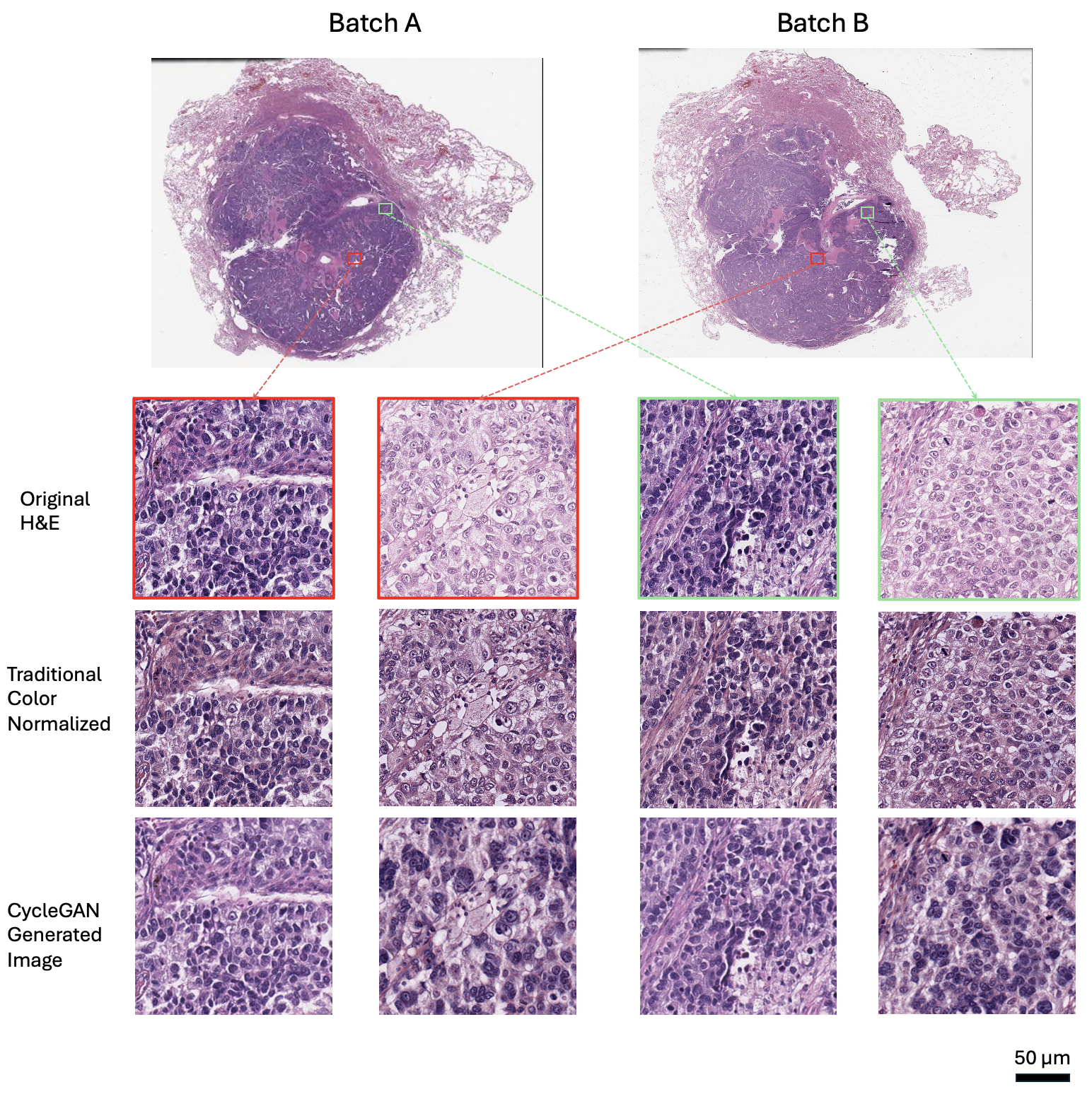}
\caption{Example images of H\&E stained tissues from adjacent cuts as well as the results of traditional and generative color normalization methods. Red and green boxes indicates two pairs of images from neighboring tissue regions. }
\label{fig:Fig5}
\end{figure}

\section*{Discussion}
We have demonstrated that when histologic slides are prepared consistently, in a short timeframe within the same lab, using the same batches of staining reagents and under similar conditions (e.g., temperature, humidity), DNN can be trained to predict subsequent metastasis based on an assessment of the original diagnostic histology. We have further shown that this process is generalizable, in that slides prepared at two different times can be used to train the DNN to perform this task. However, we have also shown that when slides processed at one time are used for training, the subsequent predictive algorithm is unable to predict metastasis on slides process at a different time. Thus, while there is clearly information present in the diagnostic histology that allows for prediction, it cannot be generalized to slides processed at a different time, even in the same lab. In our study, the slides processed at two different times in the same lab have tinctorial differences that a human pathologist can appreciate, which we suspect is the root cause of the lack of generalizability between batches. However, we have now also demonstrated that use of two different color normalization schemes is unable to make the DNN algorithms more generalizable and have further noted that with both normalization schemes we used, traditional structure-preserving method and CycleGAN-based generative method, the normalized images continue to have tinctorial and morphologic differences from the reference batch. This suggests that in order to make DNN algorithms more generalizable across histologic preparations not only across time but also location, we need to move away from vital staining, and to other ways to analyze histologic slides and images.

The implication for digital pathology and DNN is that for more common tasks where large amounts of diverse data can be easily obtained, the big data approach remains the most viable option to mitigate the effects of unwanted variations. Moving forward, there is a need for better and more sophisticated color normalization methods. This study also indicates that there is another way forward – improve the imaging and sample processing protocols to yield more consistently stained histopathology slides. For instance, it is worth considering circumventing the use of H\&E staining altogether and implement microscopy techniques to collect label-free images for downstream DNN analysis. While a human pathologist may spend years training and specializing on reading H\&E stained slides well, DNN can be readily trained on different contrasts as long as consistent data are available. The options for label-free imaging are abundant and growing in numbers. For example, Fourier Ptychographic Microscopy (FPM)\cite{zheng_wide-field_2013,horstmeyer_digital_2015} and more recent technology Angular Ptychographic Imaging with Closed-Form solution (APIC)\cite{cao_high-resolution_2024,zhao_efficient_2024} is able to provide aberration-free, quantitative phase imaging. Other label-free methods can utilize autofluorescence signals, or ultraviolet light to elicit more molecular specificity\cite{you_intravital_2018, tu_stain-free_2016}. 

\bibliography{StainVar}

\section*{Acknowledgements}

This study was supported by U01CA233363 from the National Cancer Institute (RJC) and by the Washington University in St. Louis School of Medicine Personalized Medicine Initiative (RJC). S.L., H.Z. and C.Y. are also supported by Heritage Research Institute for the Advancement of Medicine and Science at Caltech (Grant No. HMRI-15-09-01) and the Caltech Rothenberg Innovation Initiative A4188-Yang-3-A1. M.W. and R.G. are also supported 5R01CA182746 from the National Cancer Institute. 

\section*{Author contributions statement}

C.Y. and S.L. designed the experiments. S.L. conducted all experiments and analysis. S.L., H.Z., and C.Y. designed the figures. R.J.C. conducted pathology analysis. M.W. and R.G. prepared experimental data. All authors contributed to the writing and preparation of the manuscript as well as figures.

\section*{Data availability}
Part of the processed data is available at CaltechData \url{https://doi.org/10.22002/dw66e-mbs82}.




\end{document}